\documentclass[twocolumn,showpacs]{revtex4-1}
\usepackage{amsfonts,amssymb,amsmath}
\usepackage{array,dcolumn}
\usepackage{graphicx}
\usepackage{bm}

\begin{document}

\title{
Breathing multichimera states in nonlocally coupled phase oscillators
}

\author{Yusuke Suda}
\author{Koji Okuda}
\affiliation{
Division of Physics, Hokkaido University, Sapporo 060-0810, Japan
}

\date{\today}

\begin{abstract}
 Chimera states for the one-dimensional array of nonlocally coupled
 phase oscillators in the continuum limit are assumed to be stationary
 states in most studies, but a few studies report the existence of
 breathing chimera states.
 We focus on multichimera states with two coherent and incoherent
 regions, and numerically demonstrate that breathing multichimera
 states, whose global order parameter oscillates temporally, can appear.
 Moreover, we show that the system exhibits a Hopf bifurcation from a
 stationary multichimera to a breathing one by the linear stability
 analysis for the stationary multichimera.
\end{abstract}


\maketitle

\section{Introduction}
\label{sec:Intro}

Coupled oscillator systems have been studied extensively in various
scientific fields for many years~\cite{Springer.1984, Cambridge.2003}.
In particular, chimera states of coupled oscillators have recently
attracted great interest~\cite{NPCS.5.380, PhysRevE.69.036213,
PhysRevLett.93.174102, IJBC.16.21, PhysRevLett.100.144102,
PhysRevLett.101.084103, PhysRevLett.101.264103, PhysicaD.238.1569,
PhysRevLett.104.044101, PhysRevE.81.065201, Chaos.21.013112,
PhysRevLett.106.234102, PhysRevE.84.015201, PhysRevE.85.026212,
NatPhys.8.662, NatPhys.8.658, PNAS.110.10563, PhysRevLett.110.224101,
Nonlinearity.26.2469, Chaos.24.013102, IJBC.24.1440014,
PhysRevE.90.022919, PhysRevE.90.030902, Chaos.25.013106,
Nonlinearity.28.R67, SciRep.5.9883, Chaos.25.064401, Chaos.25.083104,
PhysRevE.92.060901, PhysRevE.93.012218, JPhysA.50.08LT01, CNSNS.56.1}.
Chimera states can be seen in a wide variety of systems with different
coupling topologies~\cite{PhysRevE.69.036213, PhysRevLett.104.044101,
PhysRevLett.101.084103, PhysRevLett.101.264103, PhysRevE.93.012218,
NatPhys.8.662, Chaos.24.013102, Chaos.25.064401, SciRep.5.9883} and
different kinds of constituent oscillators~\cite{NatPhys.8.662,
Chaos.24.013102, Chaos.25.064401, SciRep.5.9883, PhysRevLett.106.234102,
PhysRevE.85.026212, Chaos.25.083104, PhysRevLett.110.224101}, and are
also found experimentally~\cite{Chaos.24.013102, NatPhys.8.658,
PNAS.110.10563, PhysRevE.90.030902}.
One of the most basic models among them is the one-dimensional array of
nonlocally coupled phase oscillators~\cite{NPCS.5.380,
PhysRevLett.93.174102, IJBC.16.21, PhysRevLett.100.144102,
PhysRevE.81.065201, PhysicaD.238.1569, PhysRevE.84.015201,
Chaos.21.013112, Nonlinearity.26.2469, PhysRevE.90.022919,
IJBC.24.1440014, JPhysA.50.08LT01, Chaos.25.013106, PhysRevE.92.060901,
CNSNS.56.1}.
The essential feature of chimera states in phase oscillator systems on
one-dimensional space is the coexistence of coherent regions of
synchronized oscillators and incoherent regions of drifting oscillators.
The morphology of their coexistence depends on the coupling function
corresponding to the interaction between oscillators and the coupling
kernel function characterizing nonlocality.
In many cases, only the fundamental harmonic component is used for the
coupling function, as in the present paper.
However, it is also reported that higher harmonic components in the
coupling function are responsible for a rich variety of chimera
states~\cite{Chaos.25.013106, PhysRevE.92.060901, CNSNS.56.1}.

Chimera states for the one-dimensional array of phase oscillators in the
continuum limit $N\to\infty$, where $N$ is the number of oscillators,
are mostly assumed to be stationary states.
This means that the local mean field or the local order parameter is
stationary in the rotating frame with a constant frequency.
This assumption about chimera states plays an important role in various
studies, e.g., the self-consistency equation of the local mean
field~\cite{NPCS.5.380, PhysRevLett.93.174102, IJBC.16.21,
PhysRevLett.100.144102} and the linear stability analysis of chimera
states~\cite{Nonlinearity.26.2469, PhysRevE.90.022919,
JPhysA.50.08LT01}.
However, Abrams {\it et al.}~\cite{PhysRevLett.101.084103} discovered
breathing chimeras, whose global order parameter of a population
oscillates temporally, for two interacting populations of globally
coupled phase oscillators, and posed the question of whether such
breathing chimeras exist in the case of one-dimensional arrays.
To answer this question, Laing~\cite{PhysicaD.238.1569} demonstrated
that there also appear breathing chimeras in the one-dimensional system
by introducing phase lag parameter heterogeneity.

In this paper, we focus on chimera states, especially multichimera
states with two coherent and incoherent regions, in one-dimensional
nonlocally coupled phase oscillators.
Moreover, it is demonstrated that breathing chimeras can appear even in
homogeneous systems without introducing parameter heterogeneity.
By numerical simulations, we observe that the appearance of breathing
chimeras depends on the coupling kernel function.
Then we show that the system exhibits a Hopf bifurcation from a
stationary chimera to a breathing one by the linear stability analysis
for the stationary chimera.

\section{Model}
\label{sec:Model}

We consider the system of nonlocally coupled phase oscillators obeying
\begin{equation}
 \dot{\theta}(x,t) = \omega - \int^{\pi}_{-\pi} dy \, G(x-y)
  \sin[\theta(x,t) - \theta(y,t) + \alpha]
 \label{eq:PhaseOsc}
\end{equation}
with $2\pi$-periodic phase $\theta(x,t)$ on the one-dimensional space
$x\in[-\pi,\pi]$ under the periodic boundary condition.
The coupling between oscillators is assumed to be the sine function with
the phase lag parameter $\alpha$~\cite{ProgTheorPhys.76.576}, and the
natural frequency $\omega$ can be set to zero without loss of
generality.
The coupling kernel function $G(x)$ is generally an even real function
described as
\begin{equation}
 G(x) = \sum_{k=0}^{\infty} \, g_k \cos(kx),
 \label{eq:Kernel}
\end{equation}
where $g_k\in\mathbb{R}$ and $x\in[-\pi,\pi]$.
Nonlocal coupling is characterized by this kernel, which can be taken
as, e.g., the exponential kernel~\cite{NPCS.5.380,
PhysRevLett.100.144102, CNSNS.56.1, JPhysA.50.08LT01} and the cosine
kernel~\cite{PhysRevLett.93.174102, PhysicaD.238.1569, IJBC.16.21,
PhysRevE.90.022919}.
In this paper, we particularly use the step
kernel~\cite{PhysRevE.81.065201, Chaos.21.013112, PhysRevE.84.015201,
Nonlinearity.26.2469, IJBC.24.1440014, PhysRevE.92.060901},
\begin{equation}
 G(x) =
  \begin{cases}
   {\displaystyle \frac{1}{2\pi r}} & (|x| \leq \pi r) \\
   \\
   \;\; 0 & (|x| > \pi r)
  \end{cases}
 \label{eq:StepKernel}
\end{equation}
with $0<r\leq1$, where $r$ denotes the coupling range.
For numerical simulation of Eq.~(\ref{eq:PhaseOsc}), we need to
discretize $x$ into $x_j:=-\pi+2\pi j/N$ ($j=0,\cdots,N-1$).
Then Eq.~(\ref{eq:PhaseOsc}) with the step kernel given by
Eq.~(\ref{eq:StepKernel}) is rewritten as
\begin{equation}
 \dot{\theta}_j(t) = \omega - \frac{1}{2R}
  \sum_{k=j-R}^{j+R} \sin[\theta_j(t) - \theta_k(t) + \alpha],
 \label{eq:DisPhaseOsc}
\end{equation}
where $\theta_j(t):=\theta(x_j,t)$ and $R:=rN/2$.
All the indexes in Eq.~(\ref{eq:DisPhaseOsc}) are regarded as modulo
$N$, considering the periodic boundary condition.
In this paper, we use the fourth-order Runge-Kutta method with time
interval $\Delta t=0.01$ for all numerical simulations.
Note that chimera states for Eq.~(\ref{eq:DisPhaseOsc}) are merely
transient for small $N$, but, when $N$ is larger, the transient time
becomes longer and diverges to infinity in the continuum limit, where
the chimera state appears stable~\cite{Chaos.21.013112,
PhysRevE.84.015201, PhysRevE.90.030902, PhysRevE.92.060901}.

\begin{figure}[tb]
 \centering
 \includegraphics[width=86truemm]{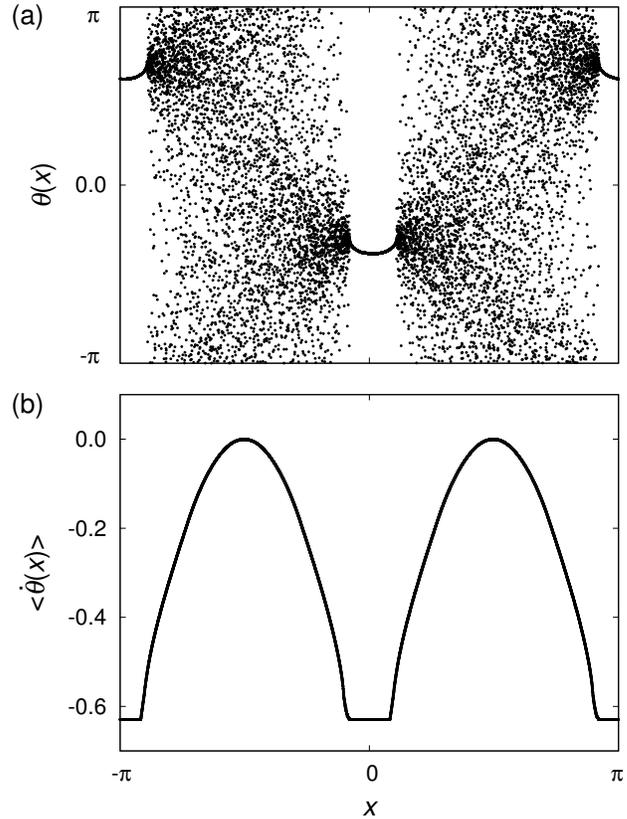}
 \caption{
 Multichimera state with two coherent and incoherent regions for
 Eq.~(\ref{eq:DisPhaseOsc}) with $N=10000$, $\alpha=1.480$, and
 $r=0.360$.
 (a)~The snapshot of the phase $\theta(x,t)$.
 (b)~The profile of the average frequency
 $\langle\dot{\theta}(x)\rangle$ with $T=2000$.
 }
 \label{fig:Chimera}
\end{figure}

\begin{figure}[tb]
 \centering
 \includegraphics[width=86truemm]{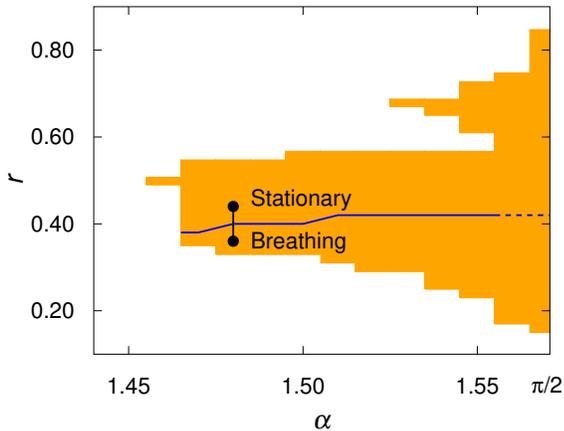}
 \caption{
 Stability region of 2-multichimera for Eq.~(\ref{eq:DisPhaseOsc})
 obtained by the numerical simulation with $N=100000$.
 Black circles denote the parameter values of
 Fig.~\ref{fig:GlobalOrder}.
 The blue line denotes the Hopf bifurcation points obtained by the
 linear stability analysis for the stationary 2-multichimera with fixed
 $\alpha$ in Sec.~\ref{sec:Breathing}, but we could not determine those
 points for $\alpha$ close to $\pi/2$ (dashed line).
 }
 \label{fig:StabRegion}
\end{figure}

For Eq.~(\ref{eq:DisPhaseOsc}), there appear various types of chimera
states~\cite{IJBC.24.1440014}, including multichimera states with two or
more coherent and incoherent regions.
A typical multichimera state obtained by numerical simulation is shown
in Fig.~\ref{fig:Chimera}.
This multichimera state has two coherent and incoherent regions, which
we call 2-multichimera below.
Figure~\ref{fig:Chimera}(a) shows the snapshot of the phase
$\theta(x,t)$.
Two coherent regions are separated from each other by the phase almost
exactly $\pi$, which is a remarkable feature of the 2-multichimera and
different from merely two neighboring chimeras.
In order to assist the emergence of 2-multichimera for the numerical
simulations of Eq.~(\ref{eq:DisPhaseOsc}), we use the following initial
condition close to a 2-multichimera~\cite{IJBC.16.21,
PhysRevE.92.060901}:
\begin{equation}
 \theta(x) \! = \!
  \begin{cases}
   {\displaystyle
   \exp
   \! \left[ -30 \left( \frac{|x|}{2\pi}-\frac{1}{4} \right)^2 \right]
   \! p(x)
   } & ( 0 \leq |x| \leq {\displaystyle \frac{\pi}{2}} ) \\
   \\
   {\displaystyle
   \exp
   \! \left[ -30 \left( \frac{|x|}{2\pi}-\frac{1}{4} \right)^2 \right]
   \! p(x) + \pi
   } & ( {\displaystyle \frac{\pi}{2}} < |x| \leq \pi ),
  \end{cases}
 \label{eq:IniPhase}
\end{equation}
where $p(x)\in[-\pi,\pi]$ is a uniform random number.
Figure~\ref{fig:Chimera}(b) shows the profile of the average frequency
$\langle\dot{\theta}(x)\rangle:=\int_{0}^{T}dt'\,\dot{\theta}(x,t')/T$
with the measurement time $T$.
When we refer to time-averaged quantities $\langle\cdot\rangle$, we set
the measurement time to $T=2000$ and measured those quantities after the
transient time 2000.
The stability region of the 2-multichimera for
Eq.~(\ref{eq:DisPhaseOsc}) obtained by the numerical simulation with
$N=100000$ is shown in Fig.~\ref{fig:StabRegion}.
This result is consistent with the phase diagram
in~\cite{IJBC.24.1440014} as far as the stability region of
2-multichimera is concerned.
However, in~\cite{IJBC.24.1440014}, the stationary and breathing
2-multichimeras are not distinguished, and the region of the breathing
one is identified as a part of the region of the stationary one.

\begin{figure}[tb]
 \centering
 \includegraphics[width=86truemm]{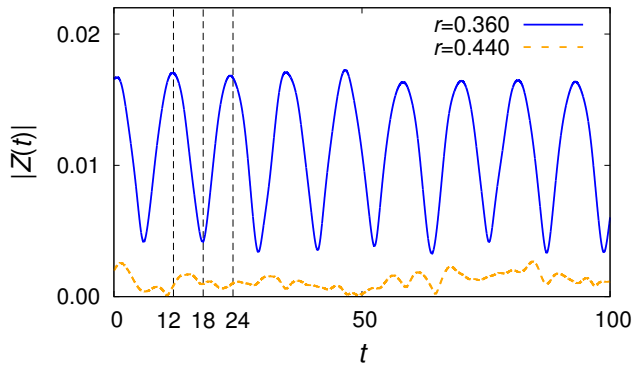}
 \caption{
 Time evolution of the global order parameter $|Z(t)|$ for a
 2-multichimera for Eq.~(\ref{eq:DisPhaseOsc}) with $N=100000$ and
 $\alpha=1.480$.
 The 2-multichimera is breathing for $r=0.360$ (blue solid line), while
 it is stationary for $r=0.440$ (orange dashed line).
 Vertical dashed lines correspond to the times $t$ in
 Fig.~\ref{fig:BreathMF}.
 }
 \label{fig:GlobalOrder}
\end{figure}

It is mostly assumed that the chimera state for Eq.~(\ref{eq:PhaseOsc})
is a stationary state in the rotating frame with a frequency $\Omega$.
This precisely means that the local mean field,
\begin{equation}
 Y(x,t) := \int^{\pi}_{-\pi} dy \, G(x-y) \, e^{i\theta(y,t)},
 \label{eq:MeanField}
\end{equation}
acting on the oscillator located in point $x$, takes the form
$Y(x,t)=\tilde{Y}(x)\,e^{i\Omega t}$.
Then the global order parameter,
\begin{equation}
 Z(t) := \frac{1}{2\pi} \int^{\pi}_{-\pi} dy \, e^{i\theta(y,t)},
 \label{eq:GlobalOrder}
\end{equation}
also takes the form $Z(t)=\tilde{Z}\,e^{i\Omega t}$.
Here, $|Z(t)|$ denotes the synchronization degree of all oscillators,
that is, all oscillators are completely synchronized in phase for
$|Z(t)|=1$ and otherwise for $0\leq|Z(t)|<1$.
In the case of the stationary 2-multichimera as in
Fig.~\ref{fig:Chimera}, $|Z(t)|$ should vanish in the continuum limit
$N\to\infty$, but we found that $|Z(t)|$ can oscillate periodically at
appropriate parameters $(\alpha ,r)$ and sufficiently large $N$.
Figure~\ref{fig:GlobalOrder} shows the time evolution of $|Z(t)|$ for
2-multichimeras with $N=100000$.
The blue solid line ($\alpha=1.480$ and $r=0.360$) exhibits a clear
periodic oscillation, while the orange dashed line ($\alpha=1.480$ and
$r=0.440$) merely exhibits a small fluctuation around zero.
We call the former state breathing 2-multichimera, while we regard the
latter as stationary 2-multichimera.

\begin{figure}[tb]
 \centering
 \includegraphics[width=86truemm]{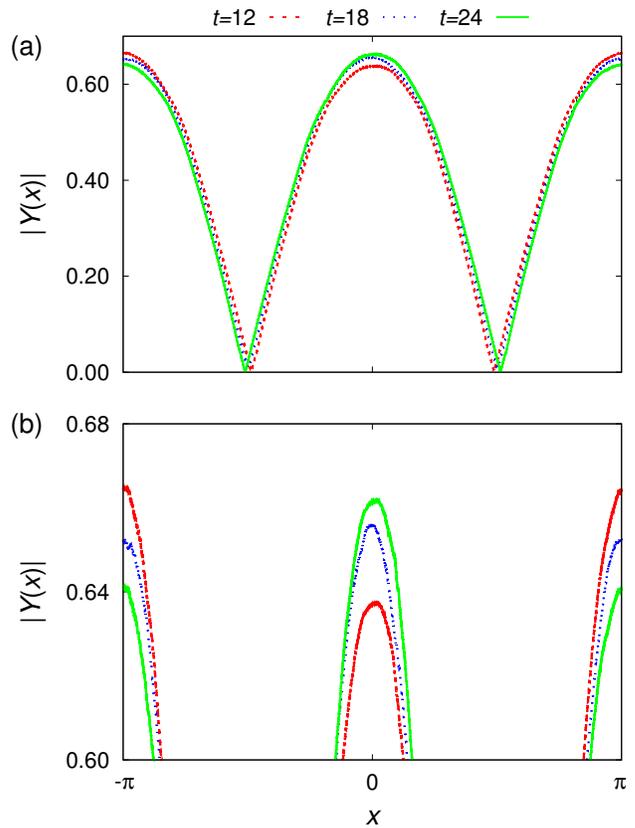}
 \caption{
 Snapshot of the local mean field $|Y(x,t)|$ for the breathing
 2-multichimera for Eq.~(\ref{eq:DisPhaseOsc}) with $N=100000$,
 $\alpha=1.480$, and $r=0.360$.
 (a)~The global view of the snapshot; (b)~the upper enlarged view.
 The red dashed line ($t=12$) and green solid line ($t=24$)
 approximately correspond to peaks of the global order parameter
 $|Z(t)|$, while the blue dotted line ($t=18$) approximately corresponds
 to a valley, as shown in Fig.~\ref{fig:GlobalOrder}.
 }
 \label{fig:BreathMF}
\end{figure}

The detailed periodic behavior of the breathing 2-multichimera can be
confirmed as the periodic oscillation of $|Y(x,t)|$ as shown in
Fig.~\ref{fig:BreathMF}.
$|Y(x,t)|$ takes a bimodal form, where the positions of the peaks
correspond to each center of the coherent regions.
Within a period of the global order parameter $|Z(t)|$ approximately
corresponding to $t=12\sim24$ in Fig.~\ref{fig:GlobalOrder}, $|Y(x,t)|$
experiences the variation in one-half of its period, and within the next
period of $|Z(t)|$, $|Y(x,t)|$ completes its whole period.
Therefore, the period of $|Y(x,t)|$ is double that of $|Z(t)|$.
In the simulation of Fig.~\ref{fig:BreathMF}, the angular frequency of
$|Y(x,t)|$ is calculated as about 0.270, which is compared with the
result of the linear stability analysis in Sec.~\ref{sec:Breathing}.

We can distinguish between stationary and breathing 2-multichimeras by
studying the time evolution of $|Z(t)|$, but that is difficult for small
$N$ because large fluctuation in $|Z(t)|$ is unavoidable.
To distinguish between these clearly, we needed 10000 oscillators at
least in our numerical simulation.
Though such breathing properties of the standard chimera states in
one-dimensional phase oscillators systems is observed by introducing
phase lag parameter heterogeneity~\cite{PhysicaD.238.1569}, we note that
the present breathing 2-multichimera does not require such
heterogeneity.
In this paper, we focus on this breathing 2-multichimera, and study the
bifurcation mechanism from the stationary 2-multichimera.

\section{Stationary 2-multichimera}
\label{sec:Stationary}

Here, we study the basic properties of stationary 2-multichimeras.
We first rewrite Eq.~(\ref{eq:PhaseOsc}) as
\begin{equation}
 \dot{\theta}(x,t) = \omega
  + {\rm Im}[e^{-i\theta(x,t)}e^{-i\alpha}Y(x,t)].
 \label{eq:PhaseOsc-2}
\end{equation}
Furthermore, we define the local order
parameter~\cite{PhysRevE.90.022919},
\begin{equation}
 z(x,t) := \lim_{\delta\to0+} \frac{1}{2\delta}
  \int^{x+\delta}_{x-\delta} dy \, e^{i\theta(y,t)},
 \label{eq:LocalOrder}
\end{equation}
and obtain $Y(x,t)=\int^{\pi}_{-\pi}dy\,G(x-y)\,z(y,t)$.
$|z(x,t)|$ denotes the synchronization degree of oscillators around
point $x$, similarly to the global order parameter $|Z(t)|$.
For $|z(x,t)|=1$, the oscillators in the neighborhood of $x$ are
completely synchronized in phase.
Otherwise, when their phases are scattered, we obtain $0\leq|z(x,t)|<1$.
Therefore, we can identify $|z(x,t)|=1$ and $0\leq|z(x,t)|<1$ as the
coherent and incoherent regions for chimera states, respectively.

Following the method in~\cite{PhysRevLett.101.264103, Chaos.21.013112}
using the Watanabe-Strogatz
approach~\cite{PhysicaD.74.197}, we can obtain the evolution equation of
$z(x,t)$ as
\begin{equation}
 \dot{z}(x,t) = i \omega z(x,t)
  + \frac{1}{2} e^{-i\alpha} Y(x,t)
  - \frac{1}{2} e^{i\alpha} z^2(x,t) Y^*(x,t),
 \label{eq:z-Eq}
\end{equation}
using Eq.~(\ref{eq:PhaseOsc-2}), where the symbol $*$ denotes the
complex conjugate.
Equation~(\ref{eq:z-Eq}) can also be obtained by another
method~\cite{PhysicaD.238.1569, Nonlinearity.26.2469} using the
Ott-Antonsen ansatz~\cite{Chaos.18.037113, Chaos.19.023117}.
Assuming the stationary solution $z(x,t)=\tilde{z}(x)\,e^{i\Omega t}$,
Eq.~(\ref{eq:z-Eq}) is rewritten as
\begin{equation}
 0 = i \Delta \tilde{z}(x)
  + \frac{1}{2} e^{-i\alpha} \tilde{Y}(x)
  - \frac{1}{2} e^{i\alpha} \tilde{z}^2(x) \tilde{Y}^*(x),
 \label{eq:Stationary-z-Eq}
\end{equation}
where $\Delta:=\omega-\Omega$.
Solving Eq.~(\ref{eq:Stationary-z-Eq}) as a quadratic equation in terms
of $\tilde{z}(x)$ and integrating the solution, we can obtain the
self-consistency equation,
\begin{equation}
 \tilde{Y}(x) = i e^{-i\alpha} \int^{\pi}_{-\pi} dy \, G(x-y) \,
  \tilde{Y}(y) \, h(y),
 \label{eq:SelfCons}
\end{equation}
\begin{equation}
 h(x) :=
  \begin{cases}
   {\displaystyle \frac{\Delta - \sqrt{\Delta^2 - R(x)^2}}{R(x)^2}}
    & [\Delta > R(x)]\\
   \\
   {\displaystyle \frac{\Delta - i\sqrt{R(x)^2 - \Delta^2}}{R(x)^2}}
    & [\Delta \leq R(x)],
  \end{cases}
 \label{eq:SelfCons-h}
\end{equation}
where $R(x)\,e^{i\Theta(x)}:=\tilde{Y}(x)=Y(x,t)\,e^{-i\Omega t}$.
These equations correspond to the self-consistency equation derived by
Kuramoto and Battogtokh~\cite{NPCS.5.380}.
Equations~(\ref{eq:SelfCons}) and~(\ref{eq:SelfCons-h}) are composed of
two equations given by the real and imaginary parts, but have three real
unknowns $R(x)$, $\Theta(x)$, and $\Delta$.
Therefore, we need to add the third condition to solve them.
The third condition can be obtained from the fact that
Eqs.~(\ref{eq:SelfCons}) and~(\ref{eq:SelfCons-h}) are invariant under
any rotation,
$\Theta(x)\to\Theta(x)+\Theta_0$~\cite{PhysRevLett.93.174102,
IJBC.16.21, PhysRevLett.100.144102, PhysRevE.90.022919}.
Based on the above, we have chosen the condition
\begin{equation}
 \Theta(-\pi) = -\frac{\pi}{2}.
 \label{eq:3rd-Cond}
\end{equation}
Equations~(\ref{eq:SelfCons}) and~(\ref{eq:SelfCons-h}) under
Eq.~(\ref{eq:3rd-Cond}) can be numerically solved by the following
iteration procedure~\cite{IJBC.16.21, PhysRevLett.100.144102}.
First, we prepare an initial function $\tilde{Y}(x)$, i.e., $R(x)$ and
$\Theta(x)$, and obtain $\Delta$ satisfying Eq.~(\ref{eq:3rd-Cond}) from
Eqs.~(\ref{eq:SelfCons}) and~(\ref{eq:SelfCons-h}) by Newton's method
with respect to $\Delta$.
Second, substituting $\tilde{Y}(x)$ and $\Delta$ into the right-hand
side of Eq.~(\ref{eq:SelfCons}), we generate a new $\tilde{Y}(x)$ from
the left-hand side.
Third, we obtain a new $\Delta$ satisfying Eq.~(\ref{eq:3rd-Cond}),
again by Newton's method, using the new $\tilde{Y}(x)$.
It only remains to repeat the second and third steps until both
$\tilde{Y}(x)$ and $\Delta$ converge.
Note that space translational symmetry of $\tilde{Y}(x)$ is not
eliminated in this iteration procedure, so the spatial position of
$\tilde{Y}(x)$ depends on the initial condition.

According to~\cite{Nonlinearity.26.2469}, it is analytically proved that
2-multichimeras exist under the condition $g_1\neq0$ for the coupling
kernel given by Eq.~(\ref{eq:Kernel}), and the local mean field
$\tilde{Y}(x)$ of a stationary 2-multichimera is given by an even
function,
\begin{equation}
 \tilde{Y}(x) = \sum_{m=1}^{\infty} C_{2m-1} \cos[(2m-1)x],
 \label{eq:MF-2multi}
\end{equation}
where $C_{2m-1} \in \mathbb{C}$.
This means that all 2-multichimera solutions of Eqs.~(\ref{eq:SelfCons})
and~(\ref{eq:SelfCons-h}) can be transformed into the form given by
Eq.~(\ref{eq:MF-2multi}) by the appropriate spatial translation.
Using Eq.~(\ref{eq:MF-2multi}), $h(x)$ turns out to be an even function
because $R(x)=|\tilde{Y}(x)|$ is also even.
Let a set of $\tilde{Y}(x)$ satisfying Eq.~(\ref{eq:MF-2multi}) and
$\Delta$ be a solution of Eqs.~(\ref{eq:SelfCons})
and~(\ref{eq:SelfCons-h}).
By substituting Eqs.~(\ref{eq:Kernel}) and~(\ref{eq:MF-2multi}) into
Eqs.~(\ref{eq:SelfCons}) and~(\ref{eq:SelfCons-h}) and eliminating the
terms whose integrands are odd functions of $y$, we obtain
\begin{eqnarray}
 \tilde{Y}(x)
 &=& 2i e^{-i\alpha}
  {\displaystyle \sum_{k=0}^{\infty}} \, g_{k} \cos(kx)
  {\displaystyle \sum_{m=1}^{\infty}} \, C_{2m-1} \nonumber \\
 &\times& {\displaystyle \int^{\pi}_{0}} dy \, \cos(ky) \, \cos[(2m-1)y]
  \, h(y).
 \label{eq:SC-Cal1}
\end{eqnarray}
Changing the variable as $y'=y-\pi/2$ in the integration in
Eq.~(\ref{eq:SC-Cal1}), the function $h(y'+\pi/2)$ in the integrand is
an even function of $y'$ because of Eq.~(\ref{eq:MF-2multi}).
We again eliminate the terms whose integrands are odd functions of $y'$,
and obtain
\begin{eqnarray}
 &&\tilde{Y}(x) = 2i e^{-i\alpha}
  {\displaystyle \sum_{l=1}^{\infty}} \, g_{2l-1} \cos[(2l-1)x]
  {\displaystyle \sum_{m=1}^{\infty}} \, C_{2m-1} (-1)^{l+m}
  \nonumber \\
 &&\times {\displaystyle \int^{\frac{\pi}{2}}_{-\frac{\pi}{2}}} dy'
  \sin[(2l-1)y'] \sin[(2m-1)y'] h(y'+\frac{\pi}{2}).
 \label{eq:SC-Cal2}
\end{eqnarray}
From the above, for $\tilde{Y}(x)$ and $\Delta$ of a stationary
2-multichimera satisfying Eqs.~(\ref{eq:SelfCons})
and~(\ref{eq:SelfCons-h}), we can finally obtain
\begin{equation}
 \tilde{Y}(x) = ie^{-i\alpha}
  \sum_{l=1}^{\infty} g_{2l-1}
  \sum_{m=1}^{\infty} C_{2m-1} A_{l m} \cos[(2l-1)x],
 \label{eq:SC-MF-2multi}
\end{equation}
where $A_{l m}$ is a complex constant.
Equation~(\ref{eq:SC-MF-2multi}) shows that $\tilde{Y}(x)$ and $\Delta$
of a stationary 2-multichimera depends only on the odd harmonic
coefficients $g_{2m-1}$, not on the even harmonic coefficients $g_{2m}$
of the coupling kernel $G(x)$.
This is because we recover Eq.~(\ref{eq:SC-MF-2multi}) even when we
substitute the identical set of $\tilde{Y}(x)$ and $\Delta$ into
Eqs.~(\ref{eq:SelfCons}) and~(\ref{eq:SelfCons-h}) with another coupling
kernel, for example,
\begin{equation}
 G_{\rm odd}(x) = \sum_{m=1}^{\infty} \, g_{2m-1} \cos[(2m-1)x],
 \label{eq:OddKernel}
\end{equation}
having the same set of odd harmonic coefficients $g_{2m-1}$.
Therefore, each stationary 2-multichimera for $G(x)$ and
$G_{\rm odd}(x)$ systems has an identical local mean field.
This does not mean that the stability properties of these
2-multichimeras are also identical.
However, Eqs.~(\ref{eq:SelfCons}) and~(\ref{eq:SelfCons-h}) for each
system have an identical solution of the stationary 2-multichimera,
whether or not each chimera is stable.
This property of the coupling kernel $G_{\rm odd}(x)$ is useful in the
linear stability analysis for the stationary 2-multichimeras mentioned
in Sec.~\ref{sec:Breathing}.

\begin{figure}[tb]
 \centering
 \includegraphics[width=86truemm]{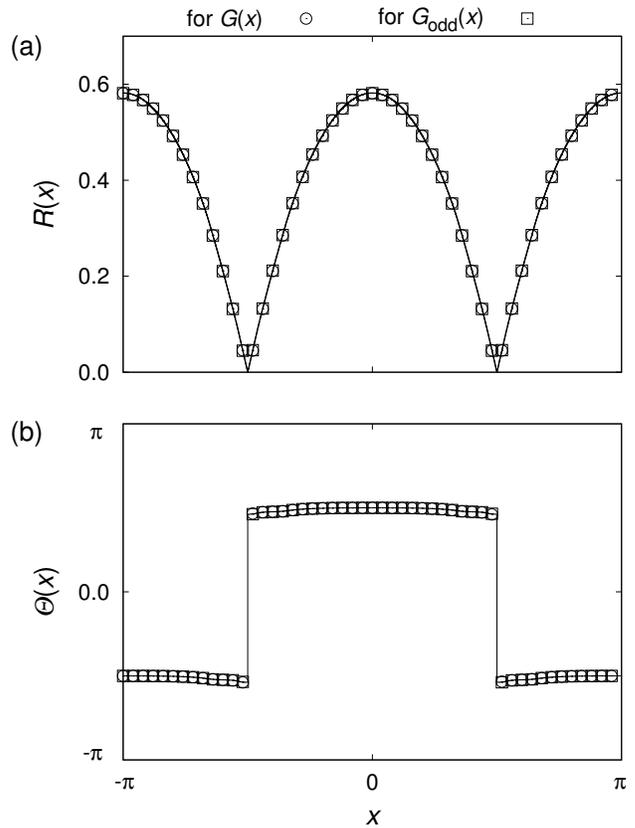}
 \caption{
 Local mean field $\tilde{Y}(x)$ of a stationary 2-multichimera.
 (a)~The amplitude $R(x)$; (b)~the argument $\Theta(x)$.
 Open circles denote the time-averaged local mean field
 $\langle\tilde{Y}(x)\rangle$ for Eq.~(\ref{eq:PhaseOsc}) with the step
 kernel $G(x)$, namely, Eq.~(\ref{eq:DisPhaseOsc}), with $N=100000$,
 $\alpha=1.480$, and $r=0.440$, and open squares denote
 $\langle\tilde{Y}(x)\rangle$ for $G_{\rm odd}(x)$ with the same
 parameters.
 Note that those are plotted once every 2000 oscillators.
 The solid line denotes the numerical solution $\tilde{Y}(x)$ to the
 self-consistency equations given by Eqs.~(\ref{eq:SelfCons})
 and~(\ref{eq:SelfCons-h}) with $G_{\rm odd}(x)$.
 }
 \label{fig:MeanField}
\end{figure}

To illustrate the above property for our step kernel given by
Eq.~(\ref{eq:StepKernel}), we performed a numerical simulation using a
new coupling kernel $G_{\rm odd}(x)$ with the same set of odd harmonic
coefficients $g_{2m-1}$ as for Eq.~(\ref{eq:StepKernel}).
By regarding Eq.~(\ref{eq:StepKernel}) as a $2\pi$-periodic function of
$x$, $G_{\rm odd}(x)$ can also be obtained as $G_{\rm
odd}(x)=[G(x)-G(x-\pi)]/2$, which we used in the numerical simulation
instead of the Fourier expansion in Eq.~(\ref{eq:OddKernel}).
Then, also for this $G_{\rm odd}(x)$, we observed stationary
2-multichimeras.
Figure~\ref{fig:MeanField} shows the time-averaged local mean fields
$\langle\tilde{Y}(x)\rangle$ of the stationary 2-multichimeras using the
step kernel $G(x)$ and the corresponding $G_{\rm odd}(x)$.
They are clearly identical, and also agree with the numerical solution
$\tilde{Y}(x)$ to the self-consistency equations given by
Eqs.~(\ref{eq:SelfCons}) and~(\ref{eq:SelfCons-h}) with
$G_{\rm odd}(x)$.
Note that the solid lines in Fig.~\ref{fig:MeanField} are obtained for
$G_{\rm odd}(x)$, not for $G(x)$.
We tried to numerically solve Eqs.~(\ref{eq:SelfCons})
and~(\ref{eq:SelfCons-h}) with the step kernel $G(x)$, but we could not
obtain a stationary solution $\tilde{Y}(x)$ of 2-multichimera because
$\tilde{Y}(x)$ converged to another solution corresponding to a standard
chimera state with one coherent and one incoherent region, under any
initial conditions.
Note that since where the iteration converges is attributed to the
property of the iteration procedure, this does not mean that the
stationary 2-multichimera solution does not exist for the $G(x)$ system.

\begin{figure}[tb]
 \centering
 \includegraphics[width=86truemm]{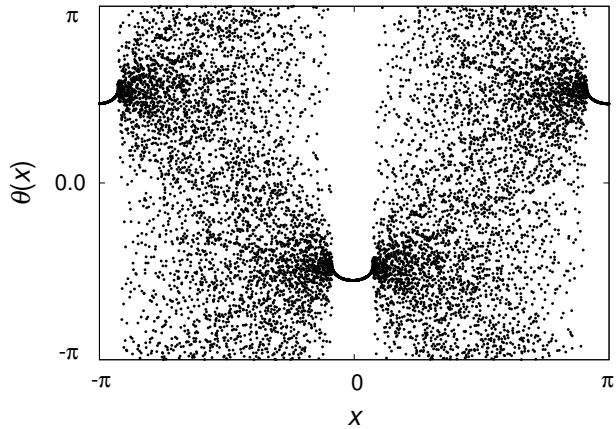}
 \caption{
 Snapshot of a 2-multichimera for Eq.~(\ref{eq:PhaseOsc}) with
 $G_{\rm odd}(x)=g_{1}\cos(x)+g_{3}\cos(3x)$ with $N=10000$,
 $\alpha=1.500$, $g_{1}=1$, and $g_{3}=-0.0916$.
 The phase $\theta(x,t)$ on any point $x$ satisfies
 Eq.~(\ref{eq:OddChimera}).
 }
 \label{fig:OddKernel}
\end{figure}

2-multichimeras can also appear for Eq.~(\ref{eq:PhaseOsc}) with other
$G_{\rm odd}(x)$, e.g.,
$G_{\rm odd}(x)=g_{1}\cos(x)$~\cite{PhysRevE.90.022919},
$G_{\rm odd}(x)=g_{1}\cos(x)+g_{3}\cos(3x)$ as shown in
Fig.~\ref{fig:OddKernel}, and so on.
In our numerical simulations for various $G_{\rm odd}(x)$ systems, we
found an interesting property common to 2-multichimeras for $G_{\rm
odd}(x)$, which is an exact relationship between the phase $\theta(x,t)$
as
\begin{equation}
 |\theta(x,t) - \theta(x-\pi,t)| = \pi,
 \label{eq:OddChimera}
\end{equation}
on any point $x$.
In fact, from Eq.~(\ref{eq:PhaseOsc-2}) with any $G_{\rm odd}(x)$, we
obtain $\dot{\theta}(x,t)-\dot{\theta}(x-\pi,t)=0$ for
Eq.~(\ref{eq:OddChimera}) by using the relation $Y(x,t)=-Y(x-\pi,t)$
satisfied at any time.
This implies that Eq.~(\ref{eq:OddChimera}) can be a solution to
Eq.~(\ref{eq:PhaseOsc}) with $G_{\rm odd}(x)$ whether stable or not, but
our simulations show that the system with $G_{\rm odd}(x)$ always
converges to the solution given by Eq.~(\ref{eq:OddChimera}) under any
initial conditions.
For the kernel other than $G_{\rm odd}(x)$, this property given by
Eq.~(\ref{eq:OddChimera}) is not exact, but seems to be satisfied only
in the meaning of average, as seen in Fig.~\ref{fig:Chimera}(a).

\begin{figure}[tb]
 \centering
 \includegraphics[width=86truemm]{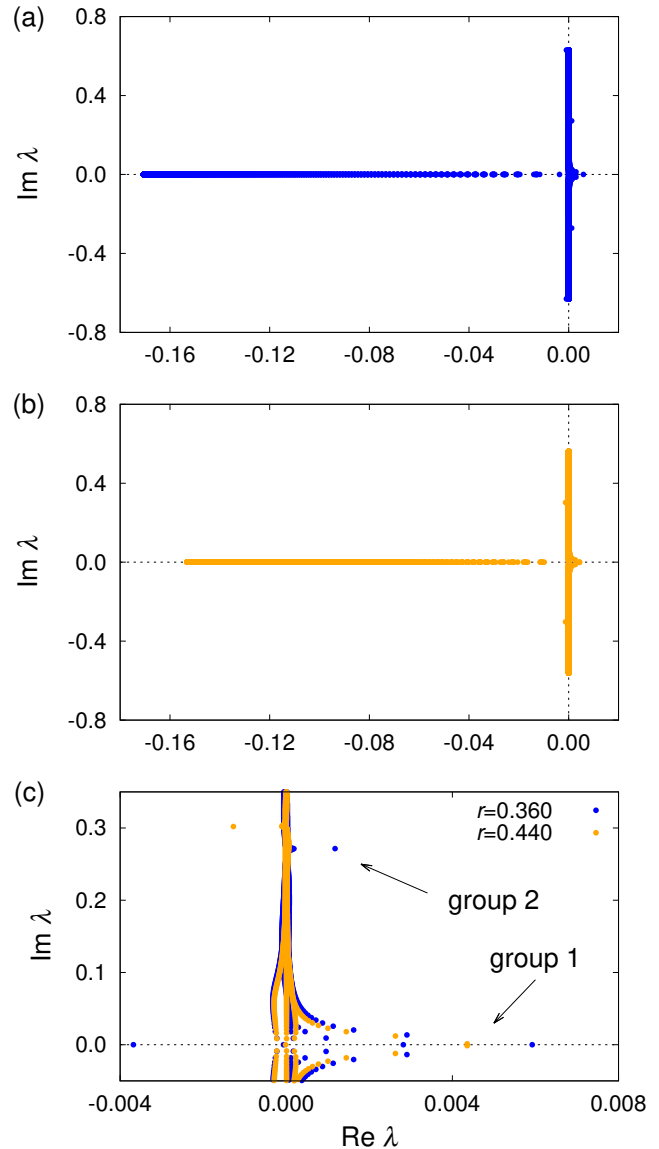}
 \caption{
 Complex eigenvalues $\lambda$ of $L_{d}$ with $M=5000$ and
 $\alpha=1.480$ using $\tilde{Y}(x)$ with $N=200000$.
 (a)~All eigenvalues for the unstable stationary 2-multichimera that
 changes to a breathing one ($r=0.360$); (b)~those for the stable
 stationary 2-multichimera ($r=0.440$).
 (c)~The enlarged view of (a) and (b) denoted by the blue and orange
 points, respectively.
 The dashed lines in each panel are drawn only for reference.
 }
 \label{fig:Eigenvalue}
\end{figure}

\section{Breathing 2-multichimera}
\label{sec:Breathing}

As described above, 2-multichimeras for Eq.~(\ref{eq:PhaseOsc}) with the
step kernel $G(x)$ do not satisfy Eq.~(\ref{eq:OddChimera}).
In our simulations, however, we often observed breathing 2-multichimeras
instead of stationary 2-multichimeras, as shown in
Fig.~\ref{fig:GlobalOrder}.
This breathing 2-multichimera is characterized by the global order
parameter $|Z(t)|$ oscillating periodically.
Therefore, 2-multichimeras for $G_{\rm odd}(x)$ satisfying
Eq.~(\ref{eq:OddChimera}) cannot breathe because the global order
parameter of Eq.~(\ref{eq:OddChimera}) exactly vanishes.

It is known that the breathing chimeras in the other
studies~\cite{PhysRevLett.101.084103, PhysicaD.238.1569,
PhysRevE.93.012218} branch via Hopf bifurcation from stable stationary
chimeras.
If the present breathing 2-multichimera also branches via Hopf
bifurcation, an unstable stationary 2-multichimera should exist in the
neighborhood of the bifurcation point.
The local mean field of this unstable stationary 2-multichimera should
be a solution to the self-consistency equations given by
Eqs.~(\ref{eq:SelfCons}) and~(\ref{eq:SelfCons-h}), and identical with
that of the stationary 2-multichimera for the $G_{\rm odd}(x)$ system.

In order to investigate this bifurcation, we analyze the linear
stability of stationary 2-multichimeras.
By substituting $z(x,t)=(\tilde{z}(x)+v(x,t))\,e^{i\Omega t}$ with the
stationary solution $\tilde{z}(x)\,e^{i\Omega t}$ and a small
perturbation $v(x,t)$ into Eq.~(\ref{eq:z-Eq}), we obtain a linear
evolution equation for $v(x,t)$,
\begin{equation}
 \dot{v}(x,t) = g(x) \tilde{z}(x)
  + \frac{1}{2} e^{-i\alpha} V(x,t)
  - \frac{1}{2} e^{i\alpha} \tilde{z}^2(x) V^*(x,t),
 \label{eq:v-Eq}
\end{equation}
\begin{equation}
 g(x) :=
  \begin{cases}
   i\sqrt{\Delta^2 - R(x)^2} & [\Delta > R(x)]\\
   \\
   -\sqrt{R(x)^2 - \Delta^2} & [\Delta \leq R(x)],
  \end{cases}
 \label{eq:v-Eq_g}
\end{equation}
\begin{equation}
 V(x,t) := \int^{\pi}_{-\pi} dy \, G(x-y) v(y,t),
 \label{eq:v-Eq_V}
\end{equation}
where $\Delta=\omega-\Omega$.
We rewrite Eqs.~(\ref{eq:v-Eq})-(\ref{eq:v-Eq_V}) as
$\dot{{\bm v}}=L\,{\bm v}$ using
${\bm v}(x,t)=({\rm Re}\,v(x,t),\,{\rm Im}\,v(x,t))^{T}$ and solve the
eigenvalue problem of $L$.
According to~\cite{Nonlinearity.26.2469, PhysRevE.90.022919}, the
spectrum of $L$ consists of the essential spectrum and the point
spectrum.
In the present case, the essential spectrum is given by $g(x)$
consisting of pure imaginary and negative real eigenvalues, which
correspond to incoherent and coherent regions, respectively.
Therefore, the stability of stationary 2-multichimeras should be
determined only by the point spectrum.

\begin{figure}[tb]
 \centering
 \includegraphics[width=86truemm]{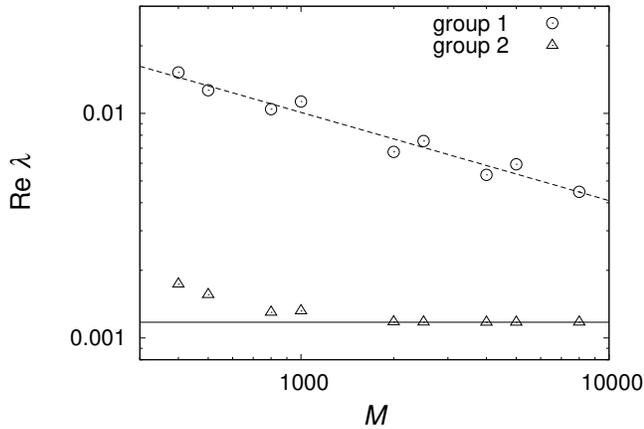}
 \caption{
 Transition of the positive real parts of the eigenvalues of $L_{d}$ for
 an unstable stationary 2-multichimera ($\alpha=1.480$ and $r=0.360$)
 with increasing $M$.
 Open circles denote the maximum values of the real parts of the
 eigenvalues in group~1, and open triangles denote those in group~2.
 The data for group~1 are fitted linearly (dashed line) in the log-log
 plot, and go to zero with increasing $M$.
 In contrast, the data for group~2 converge to a positive constant
 $1.175\times10^{-3}$ (solid line).
 }
 \label{fig:M-vs-Eigen}
\end{figure}

If the number of nonzero $g_k$ in Eq.~(\ref{eq:Kernel}) is finite, we
may solve the eigenvalue problem of a finite size matrix to obtain the
point spectrum~\cite{Nonlinearity.26.2469, PhysRevE.90.022919}.
However, the step kernel given by Eq.~(\ref{eq:StepKernel}) has infinite
numbers of nonzero $g_k$.
Therefore, we discretize the space coordinate $x\to x_{j}=-\pi+2\pi j/M$
($j=0,\cdots,M-1$) and compute all eigenvalues
$\lambda$ by solving the eigenvalue problem of $2M\times2M$ matrix
$L_{d}$ such that $\dot{{\bm v}}_{d}=L_{d}\,{\bm v}_{d}$ with
${\bm v}_{d}(t)=
(\cdots,\,{\rm Re}\,v(x_{j},t),\,{\rm Im}\,v(x_{j},t),\,\cdots)^{T}$
~\cite{JPhysA.50.08LT01, SciRep.7.2104}.
In order to solve this problem, we first need to prepare $\tilde{Y}(x)$
and $\Delta$ of the stationary 2-multichimera for
Eq.~(\ref{eq:DisPhaseOsc}) numerically, but could not obtain them by
solving the self-consistency equations given by Eqs.~(\ref{eq:SelfCons})
and~(\ref{eq:SelfCons-h}) with the step kernel $G(x)$ because
$\tilde{Y}(x)$ converged to another solution by the iteration procedure,
as mentioned in Sec.~\ref{sec:Stationary}.
We accordingly used $\tilde{Y}(x)$ and $\Delta$ of
Eqs.~(\ref{eq:SelfCons}) and~(\ref{eq:SelfCons-h}) with the
corresponding $G_{\rm odd}(x)$, instead of $G(x)$.
Although $G_{\rm odd}(x)$ is used for computing $\tilde{Y}(x)$ and
$\Delta$, we note that we insert the original kernel $G(x)$ into
Eqs.~(\ref{eq:v-Eq})-(\ref{eq:v-Eq_V}) to solve the eigenvalue problem.
Figure~\ref{fig:Eigenvalue}(a) shows all the eigenvalues $\lambda$ of
$L_{d}$ with $M=5000$, $\alpha=1.480$, and $r=0.360$ on the complex
plane.
As seen from the figure, we have some eigenvalues with positive real
part, because the stationary 2-multichimera is unstable and changes into
a breathing one at these parameters.
We can regard those eigenvalues as roughly separating into two groups.
Group~1 consists of some eigenvalues around the real axis and group~2
consists of others around the imaginary values about $0.270$ and their
complex conjugate, as shown in Fig.~\ref{fig:Eigenvalue}(c).

Even though we can observe the eigenvalues with a positive real part, we
cannot easily tell whether they belong to the point spectrum or a
fluctuation of the essential spectrum caused by finite discretization.
If an eigenvalue with a positive real part belongs to such a fluctuation,
its real part should go to zero in the continuum limit $M\to\infty$,
while an eigenvalue in the point spectrum keeps the positive real part
in that limit.
We computed the eigenvalues of $L_{d}$ with various $M$ and found their
limiting behaviors as $M$ is increased, as shown in
Fig.~\ref{fig:M-vs-Eigen}.

From Fig.~\ref{fig:M-vs-Eigen}, we can see that the maximum value of the
real parts of the eigenvalues in group~1 tends to go to zero, while that
value in group~2 converges to a positive constant.
Therefore, it turns out that at least a pair of the complex conjugate
eigenvalues in group~2 belongs to the point spectrum, while the
eigenvalues in group~1 belong to the fluctuation of the essential
spectrum by finite discretization.
At the other parameters where the stationary 2-multichimera is stable,
the point spectrum contains only the eigenvalues with a negative real
part, as shown in Fig.~\ref{fig:Eigenvalue}(b).

\begin{figure}[tb]
 \centering
 \includegraphics[width=86truemm]{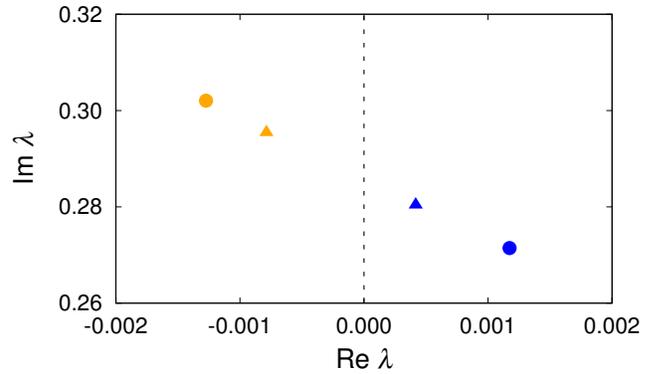}
 \caption{
 Hopf bifurcation for fixed $\alpha=1.480$.
 Each point shows the eigenvalue with the positive imaginary part and
 the maximum absolute value of the real part in the point spectrum for
 $r=0.440$ (orange circle), $r=0.420$ (orange triangle), $r=0.380$ (blue
 triangle), and $r=0.360$ (blue circle), and corresponds to the black
 solid line in Fig.~\ref{fig:StabRegion}.
 The point for $r=0.400$ is omitted in the figure because we could not
 distinguish the point spectrum from other eigenvalues for $r=0.400$
 that is very close to the Hopf bifurcation point.
 The dashed line is the imaginary axis.
 }
 \label{fig:PointSpec}
\end{figure}

Figure~\ref{fig:PointSpec} shows that a Hopf bifurcation from a
stationary 2-multichimera to a breathing one occurs for $\alpha=1.480$,
denoted by the vertical black solid line in Fig.~\ref{fig:StabRegion}.
The Hopf bifurcation points for $\alpha=1.480$ and other values are
shown as the blue line in Fig.~\ref{fig:StabRegion}.
However, we could not determine the bifurcation points for $\alpha$
close to $\pi/2$ because it is difficult to distinguish the point
spectrum whose real parts are almost zero for those $\alpha$.
We note that the absolute values of the imaginary parts of the point
spectrum, as shown in Fig.~\ref{fig:Eigenvalue}(a), are nearly equal to
the angular frequency of the local mean field $|Y(x,t)|$, as shown in
Fig.~\ref{fig:BreathMF}, which is calculated to be about 0.270.
This result agrees with the occurrence of a supercritical Hopf
bifurcation.

\section{Summary}
\label{sec:Summary}

We studied 2-multichimera with two coherent and incoherent regions in
one-dimensional nonlocally coupled phase oscillators described as
Eq.~(\ref{eq:PhaseOsc}).
First, we showed that the local mean field $\tilde{Y}(x)$ of stationary
2-multichimeras depends on only odd harmonic coefficients $g_{2m-1}$ of
the coupling kernel function $G(x)$.
This implies that if $G(x)$ has the same set of the odd harmonic
coefficients, then $\tilde{Y}(x)$ of stationary 2-multichimeras are
common to all those $G(x)$ systems for the same parameters.
We could actually apply $\tilde{Y}(x)$ of the $G_{\rm odd}(x)$ system to
the linear stability analysis of stationary 2-multichimeras for the
$G(x)$ system, even though we could not obtain $\tilde{Y}(x)$ of
stationary 2-multichimeras for the $G(x)$ system.
The method used in this paper is based on the fact that $\tilde{Y}(x)$
of stationary 2-multichimeras is characterized by only odd harmonic
components, namely, Eq.~(\ref{eq:MF-2multi}).
We expect that a similar method is applied to other stationary
multichimera states because their local mean fields are also
characterized by a set of specific harmonic
components~\cite{Nonlinearity.26.2469}.

Next, we numerically found that breathing 2-multichimeras with
oscillatory global order parameter $|Z(t)|$ appear for
Eq.~(\ref{eq:PhaseOsc}) with the step kernel given by
Eq.~(\ref{eq:StepKernel}) without introducing parameter
heterogeneity~\cite{PhysicaD.238.1569}.
Moreover, we clarified that the system exhibits a Hopf bifurcation from
a stationary 2-multichimera to a breathing one by the linear stability
analysis for the stationary 2-multichimera.
In contrast to the $G(x)$ system, 2-multichimeras for $G_{\rm odd}(x)$
cannot breathe because the system converges to the solution given by
Eq.~(\ref{eq:OddChimera}) with vanishing $|Z(t)|$.
Therefore, it is inferred that the coupling kernel is an important
factor for the appearance of breathing chimeras in the one-dimensional
system.
It may be interesting to find other breathing chimeras by using the
appropriate coupling kernel similarly, but it is an open problem.

\bibliography{Reference}

\begin{thebibliography}{39}%
\makeatletter
\providecommand \@ifxundefined [1]{%
 \@ifx{#1\undefined}
}%
\providecommand \@ifnum [1]{%
 \ifnum #1\expandafter \@firstoftwo
 \else \expandafter \@secondoftwo
 \fi
}%
\providecommand \@ifx [1]{%
 \ifx #1\expandafter \@firstoftwo
 \else \expandafter \@secondoftwo
 \fi
}%
\providecommand \natexlab [1]{#1}%
\providecommand \enquote  [1]{``#1''}%
\providecommand \bibnamefont  [1]{#1}%
\providecommand \bibfnamefont [1]{#1}%
\providecommand \citenamefont [1]{#1}%
\providecommand \href@noop [0]{\@secondoftwo}%
\providecommand \href [0]{\begingroup \@sanitize@url \@href}%
\providecommand \@href[1]{\@@startlink{#1}\@@href}%
\providecommand \@@href[1]{\endgroup#1\@@endlink}%
\providecommand \@sanitize@url [0]{\catcode `\\12\catcode `\$12\catcode
  `\&12\catcode `\#12\catcode `\^12\catcode `\_12\catcode `\%12\relax}%
\providecommand \@@startlink[1]{}%
\providecommand \@@endlink[0]{}%
\providecommand \url  [0]{\begingroup\@sanitize@url \@url }%
\providecommand \@url [1]{\endgroup\@href {#1}{\urlprefix }}%
\providecommand \urlprefix  [0]{URL }%
\providecommand \Eprint [0]{\href }%
\providecommand \doibase [0]{http://dx.doi.org/}%
\providecommand \selectlanguage [0]{\@gobble}%
\providecommand \bibinfo  [0]{\@secondoftwo}%
\providecommand \bibfield  [0]{\@secondoftwo}%
\providecommand \translation [1]{[#1]}%
\providecommand \BibitemOpen [0]{}%
\providecommand \bibitemStop [0]{}%
\providecommand \bibitemNoStop [0]{.\EOS\space}%
\providecommand \EOS [0]{\spacefactor3000\relax}%
\providecommand \BibitemShut  [1]{\csname bibitem#1\endcsname}%
\let\auto@bib@innerbib\@empty
%</preamble>
\bibitem [{\citenamefont {Kuramoto}(1984)}]{Springer.1984}%
  \BibitemOpen
  \bibfield  {author} {\bibinfo {author} {\bibfnamefont {Y.}~\bibnamefont
  {Kuramoto}},\ }\href@noop {} {\emph {\bibinfo {title} {Chemical Oscillation,
  Waves, and Turbulence}}}\ (\bibinfo  {publisher} {Springer},\ \bibinfo
  {address} {Berlin},\ \bibinfo {year} {1984})\BibitemShut {NoStop}%
\bibitem [{\citenamefont {Pikovsky}\ \emph {et~al.}(2003)\citenamefont
  {Pikovsky}, \citenamefont {Rosenblum},\ and\ \citenamefont
  {Kurths}}]{Cambridge.2003}%
  \BibitemOpen
  \bibfield  {author} {\bibinfo {author} {\bibfnamefont {A.}~\bibnamefont
  {Pikovsky}}, \bibinfo {author} {\bibfnamefont {M.}~\bibnamefont {Rosenblum}},
  \ and\ \bibinfo {author} {\bibfnamefont {J.}~\bibnamefont {Kurths}},\
  }\href@noop {} {\emph {\bibinfo {title} {Synchronization: A Universal Concept
  in Nonlinear Sciences}}}\ (\bibinfo  {publisher} {Cambridge University
  Press},\ \bibinfo {address} {Cambridge},\ \bibinfo {year} {2003})\BibitemShut
  {NoStop}%
\bibitem [{\citenamefont {Kuramoto}\ and\ \citenamefont
  {Battogtokh}(2002)}]{NPCS.5.380}%
  \BibitemOpen
  \bibfield  {author} {\bibinfo {author} {\bibfnamefont {Y.}~\bibnamefont
  {Kuramoto}}\ and\ \bibinfo {author} {\bibfnamefont {D.}~\bibnamefont
  {Battogtokh}},\ }\href@noop {} {\bibfield  {journal} {\bibinfo  {journal}
  {Nonlinear Phenom. Complex Syst.}\ }\textbf {\bibinfo {volume} {5}},\
  \bibinfo {pages} {380} (\bibinfo {year} {2002})}\BibitemShut {NoStop}%
\bibitem [{\citenamefont {Shima}\ and\ \citenamefont
  {Kuramoto}(2004)}]{PhysRevE.69.036213}%
  \BibitemOpen
  \bibfield  {author} {\bibinfo {author} {\bibfnamefont {S.-i.}\ \bibnamefont
  {Shima}}\ and\ \bibinfo {author} {\bibfnamefont {Y.}~\bibnamefont
  {Kuramoto}},\ }\href {\doibase 10.1103/PhysRevE.69.036213} {\bibfield
  {journal} {\bibinfo  {journal} {Phys. Rev. E}\ }\textbf {\bibinfo {volume}
  {69}},\ \bibinfo {pages} {036213} (\bibinfo {year} {2004})}\BibitemShut
  {NoStop}%
\bibitem [{\citenamefont {Abrams}\ and\ \citenamefont
  {Strogatz}(2004)}]{PhysRevLett.93.174102}%
  \BibitemOpen
  \bibfield  {author} {\bibinfo {author} {\bibfnamefont {D.~M.}\ \bibnamefont
  {Abrams}}\ and\ \bibinfo {author} {\bibfnamefont {S.~H.}\ \bibnamefont
  {Strogatz}},\ }\href {\doibase 10.1103/PhysRevLett.93.174102} {\bibfield
  {journal} {\bibinfo  {journal} {Phys. Rev. Lett.}\ }\textbf {\bibinfo
  {volume} {93}},\ \bibinfo {pages} {174102} (\bibinfo {year}
  {2004})}\BibitemShut {NoStop}%
\bibitem [{\citenamefont {Abrams}\ and\ \citenamefont
  {Strogatz}(2006)}]{IJBC.16.21}%
  \BibitemOpen
  \bibfield  {author} {\bibinfo {author} {\bibfnamefont {D.~M.}\ \bibnamefont
  {Abrams}}\ and\ \bibinfo {author} {\bibfnamefont {S.~H.}\ \bibnamefont
  {Strogatz}},\ }\href {\doibase 10.1142/S0218127406014551} {\bibfield
  {journal} {\bibinfo  {journal} {Int. J. Bifurc. Chaos}\ }\textbf {\bibinfo
  {volume} {16}},\ \bibinfo {pages} {21} (\bibinfo {year} {2006})}\BibitemShut
  {NoStop}%
\bibitem [{\citenamefont {Sethia}\ \emph {et~al.}(2008)\citenamefont {Sethia},
  \citenamefont {Sen},\ and\ \citenamefont {Atay}}]{PhysRevLett.100.144102}%
  \BibitemOpen
  \bibfield  {author} {\bibinfo {author} {\bibfnamefont {G.~C.}\ \bibnamefont
  {Sethia}}, \bibinfo {author} {\bibfnamefont {A.}~\bibnamefont {Sen}}, \ and\
  \bibinfo {author} {\bibfnamefont {F.~M.}\ \bibnamefont {Atay}},\ }\href
  {\doibase 10.1103/PhysRevLett.100.144102} {\bibfield  {journal} {\bibinfo
  {journal} {Phys. Rev. Lett.}\ }\textbf {\bibinfo {volume} {100}},\ \bibinfo
  {pages} {144102} (\bibinfo {year} {2008})}\BibitemShut {NoStop}%
\bibitem [{\citenamefont {Abrams}\ \emph {et~al.}(2008)\citenamefont {Abrams},
  \citenamefont {Mirollo}, \citenamefont {Strogatz},\ and\ \citenamefont
  {Wiley}}]{PhysRevLett.101.084103}%
  \BibitemOpen
  \bibfield  {author} {\bibinfo {author} {\bibfnamefont {D.~M.}\ \bibnamefont
  {Abrams}}, \bibinfo {author} {\bibfnamefont {R.}~\bibnamefont {Mirollo}},
  \bibinfo {author} {\bibfnamefont {S.~H.}\ \bibnamefont {Strogatz}}, \ and\
  \bibinfo {author} {\bibfnamefont {D.~A.}\ \bibnamefont {Wiley}},\ }\href
  {\doibase 10.1103/PhysRevLett.101.084103} {\bibfield  {journal} {\bibinfo
  {journal} {Phys. Rev. Lett.}\ }\textbf {\bibinfo {volume} {101}},\ \bibinfo
  {pages} {084103} (\bibinfo {year} {2008})}\BibitemShut {NoStop}%
\bibitem [{\citenamefont {Pikovsky}\ and\ \citenamefont
  {Rosenblum}(2008)}]{PhysRevLett.101.264103}%
  \BibitemOpen
  \bibfield  {author} {\bibinfo {author} {\bibfnamefont {A.}~\bibnamefont
  {Pikovsky}}\ and\ \bibinfo {author} {\bibfnamefont {M.}~\bibnamefont
  {Rosenblum}},\ }\href {\doibase 10.1103/PhysRevLett.101.264103} {\bibfield
  {journal} {\bibinfo  {journal} {Phys. Rev. Lett.}\ }\textbf {\bibinfo
  {volume} {101}},\ \bibinfo {pages} {264103} (\bibinfo {year}
  {2008})}\BibitemShut {NoStop}%
\bibitem [{\citenamefont {Laing}(2009)}]{PhysicaD.238.1569}%
  \BibitemOpen
  \bibfield  {author} {\bibinfo {author} {\bibfnamefont {C.~R.}\ \bibnamefont
  {Laing}},\ }\href {\doibase http://dx.doi.org/10.1016/j.physd.2009.04.012}
  {\bibfield  {journal} {\bibinfo  {journal} {Physica D}\ }\textbf {\bibinfo
  {volume} {238}},\ \bibinfo {pages} {1569 } (\bibinfo {year}
  {2009})}\BibitemShut {NoStop}%
\bibitem [{\citenamefont {Martens}\ \emph {et~al.}(2010)\citenamefont
  {Martens}, \citenamefont {Laing},\ and\ \citenamefont
  {Strogatz}}]{PhysRevLett.104.044101}%
  \BibitemOpen
  \bibfield  {author} {\bibinfo {author} {\bibfnamefont {E.~A.}\ \bibnamefont
  {Martens}}, \bibinfo {author} {\bibfnamefont {C.~R.}\ \bibnamefont {Laing}},
  \ and\ \bibinfo {author} {\bibfnamefont {S.~H.}\ \bibnamefont {Strogatz}},\
  }\href {\doibase 10.1103/PhysRevLett.104.044101} {\bibfield  {journal}
  {\bibinfo  {journal} {Phys. Rev. Lett.}\ }\textbf {\bibinfo {volume} {104}},\
  \bibinfo {pages} {044101} (\bibinfo {year} {2010})}\BibitemShut {NoStop}%
\bibitem [{\citenamefont {Omel'chenko}\ \emph {et~al.}(2010)\citenamefont
  {Omel'chenko}, \citenamefont {Wolfrum},\ and\ \citenamefont
  {Maistrenko}}]{PhysRevE.81.065201}%
  \BibitemOpen
  \bibfield  {author} {\bibinfo {author} {\bibfnamefont {O.~E.}\ \bibnamefont
  {Omel'chenko}}, \bibinfo {author} {\bibfnamefont {M.}~\bibnamefont
  {Wolfrum}}, \ and\ \bibinfo {author} {\bibfnamefont {Y.~L.}\ \bibnamefont
  {Maistrenko}},\ }\href {\doibase 10.1103/PhysRevE.81.065201} {\bibfield
  {journal} {\bibinfo  {journal} {Phys. Rev. E}\ }\textbf {\bibinfo {volume}
  {81}},\ \bibinfo {pages} {065201} (\bibinfo {year} {2010})}\BibitemShut
  {NoStop}%
\bibitem [{\citenamefont {Wolfrum}\ \emph {et~al.}(2011)\citenamefont
  {Wolfrum}, \citenamefont {Omel'chenko}, \citenamefont {Yanchuk},\ and\
  \citenamefont {Maistrenko}}]{Chaos.21.013112}%
  \BibitemOpen
  \bibfield  {author} {\bibinfo {author} {\bibfnamefont {M.}~\bibnamefont
  {Wolfrum}}, \bibinfo {author} {\bibfnamefont {O.~E.}\ \bibnamefont
  {Omel'chenko}}, \bibinfo {author} {\bibfnamefont {S.}~\bibnamefont
  {Yanchuk}}, \ and\ \bibinfo {author} {\bibfnamefont {Y.~L.}\ \bibnamefont
  {Maistrenko}},\ }\href {\doibase 10.1063/1.3563579} {\bibfield  {journal}
  {\bibinfo  {journal} {Chaos}\ }\textbf {\bibinfo {volume} {21}},\ \bibinfo
  {pages} {013112} (\bibinfo {year} {2011})}\BibitemShut {NoStop}%
\bibitem [{\citenamefont {Omelchenko}\ \emph {et~al.}(2011)\citenamefont
  {Omelchenko}, \citenamefont {Maistrenko}, \citenamefont {H\"ovel},\ and\
  \citenamefont {Sch\"oll}}]{PhysRevLett.106.234102}%
  \BibitemOpen
  \bibfield  {author} {\bibinfo {author} {\bibfnamefont {I.}~\bibnamefont
  {Omelchenko}}, \bibinfo {author} {\bibfnamefont {Y.}~\bibnamefont
  {Maistrenko}}, \bibinfo {author} {\bibfnamefont {P.}~\bibnamefont {H\"ovel}},
  \ and\ \bibinfo {author} {\bibfnamefont {E.}~\bibnamefont {Sch\"oll}},\
  }\href {\doibase 10.1103/PhysRevLett.106.234102} {\bibfield  {journal}
  {\bibinfo  {journal} {Phys. Rev. Lett.}\ }\textbf {\bibinfo {volume} {106}},\
  \bibinfo {pages} {234102} (\bibinfo {year} {2011})}\BibitemShut {NoStop}%
\bibitem [{\citenamefont {Wolfrum}\ and\ \citenamefont
  {Omel'chenko}(2011)}]{PhysRevE.84.015201}%
  \BibitemOpen
  \bibfield  {author} {\bibinfo {author} {\bibfnamefont {M.}~\bibnamefont
  {Wolfrum}}\ and\ \bibinfo {author} {\bibfnamefont {O.~E.}\ \bibnamefont
  {Omel'chenko}},\ }\href {\doibase 10.1103/PhysRevE.84.015201} {\bibfield
  {journal} {\bibinfo  {journal} {Phys. Rev. E}\ }\textbf {\bibinfo {volume}
  {84}},\ \bibinfo {pages} {015201} (\bibinfo {year} {2011})}\BibitemShut
  {NoStop}%
\bibitem [{\citenamefont {Omelchenko}\ \emph {et~al.}(2012)\citenamefont
  {Omelchenko}, \citenamefont {Riemenschneider}, \citenamefont {H\"ovel},
  \citenamefont {Maistrenko},\ and\ \citenamefont
  {Sch\"oll}}]{PhysRevE.85.026212}%
  \BibitemOpen
  \bibfield  {author} {\bibinfo {author} {\bibfnamefont {I.}~\bibnamefont
  {Omelchenko}}, \bibinfo {author} {\bibfnamefont {B.}~\bibnamefont
  {Riemenschneider}}, \bibinfo {author} {\bibfnamefont {P.}~\bibnamefont
  {H\"ovel}}, \bibinfo {author} {\bibfnamefont {Y.}~\bibnamefont {Maistrenko}},
  \ and\ \bibinfo {author} {\bibfnamefont {E.}~\bibnamefont {Sch\"oll}},\
  }\href {\doibase 10.1103/PhysRevE.85.026212} {\bibfield  {journal} {\bibinfo
  {journal} {Phys. Rev. E}\ }\textbf {\bibinfo {volume} {85}},\ \bibinfo
  {pages} {026212} (\bibinfo {year} {2012})}\BibitemShut {NoStop}%
\bibitem [{\citenamefont {Tinsley}\ \emph {et~al.}(2012)\citenamefont
  {Tinsley}, \citenamefont {Nkomo},\ and\ \citenamefont
  {Showalter}}]{NatPhys.8.662}%
  \BibitemOpen
  \bibfield  {author} {\bibinfo {author} {\bibfnamefont {M.~R.}\ \bibnamefont
  {Tinsley}}, \bibinfo {author} {\bibfnamefont {S.}~\bibnamefont {Nkomo}}, \
  and\ \bibinfo {author} {\bibfnamefont {K.}~\bibnamefont {Showalter}},\ }\href
  {\doibase 10.1038/nphys2371} {\bibfield  {journal} {\bibinfo  {journal} {Nat.
  Phys.}\ }\textbf {\bibinfo {volume} {8}},\ \bibinfo {pages} {662} (\bibinfo
  {year} {2012})}\BibitemShut {NoStop}%
\bibitem [{\citenamefont {Hagerstrom}\ \emph {et~al.}(2012)\citenamefont
  {Hagerstrom}, \citenamefont {Murphy}, \citenamefont {Roy}, \citenamefont
  {H{\"o}vel}, \citenamefont {Omelchenko},\ and\ \citenamefont
  {Sch{\"o}ll}}]{NatPhys.8.658}%
  \BibitemOpen
  \bibfield  {author} {\bibinfo {author} {\bibfnamefont {A.~M.}\ \bibnamefont
  {Hagerstrom}}, \bibinfo {author} {\bibfnamefont {T.~E.}\ \bibnamefont
  {Murphy}}, \bibinfo {author} {\bibfnamefont {R.}~\bibnamefont {Roy}},
  \bibinfo {author} {\bibfnamefont {P.}~\bibnamefont {H{\"o}vel}}, \bibinfo
  {author} {\bibfnamefont {I.}~\bibnamefont {Omelchenko}}, \ and\ \bibinfo
  {author} {\bibfnamefont {E.}~\bibnamefont {Sch{\"o}ll}},\ }\href {\doibase
  10.1038/nphys2372} {\bibfield  {journal} {\bibinfo  {journal} {Nat. Phys.}\
  }\textbf {\bibinfo {volume} {8}},\ \bibinfo {pages} {658} (\bibinfo {year}
  {2012})}\BibitemShut {NoStop}%
\bibitem [{\citenamefont {Martens}\ \emph {et~al.}(2013)\citenamefont
  {Martens}, \citenamefont {Thutupalli}, \citenamefont {Fourri{\'e}re},\ and\
  \citenamefont {Hallatschek}}]{PNAS.110.10563}%
  \BibitemOpen
  \bibfield  {author} {\bibinfo {author} {\bibfnamefont {E.~A.}\ \bibnamefont
  {Martens}}, \bibinfo {author} {\bibfnamefont {S.}~\bibnamefont {Thutupalli}},
  \bibinfo {author} {\bibfnamefont {A.}~\bibnamefont {Fourri{\'e}re}}, \ and\
  \bibinfo {author} {\bibfnamefont {O.}~\bibnamefont {Hallatschek}},\ }\href
  {\doibase 10.1073/pnas.1302880110} {\bibfield  {journal} {\bibinfo  {journal}
  {Proc. Natl. Acad. Sci. USA}\ }\textbf {\bibinfo {volume} {110}},\ \bibinfo
  {pages} {10563} (\bibinfo {year} {2013})}\BibitemShut {NoStop}%
\bibitem [{\citenamefont {Omelchenko}\ \emph {et~al.}(2013)\citenamefont
  {Omelchenko}, \citenamefont {Omel'chenko}, \citenamefont {H\"ovel},\ and\
  \citenamefont {Sch\"oll}}]{PhysRevLett.110.224101}%
  \BibitemOpen
  \bibfield  {author} {\bibinfo {author} {\bibfnamefont {I.}~\bibnamefont
  {Omelchenko}}, \bibinfo {author} {\bibfnamefont {O.~E.}\ \bibnamefont
  {Omel'chenko}}, \bibinfo {author} {\bibfnamefont {P.}~\bibnamefont
  {H\"ovel}}, \ and\ \bibinfo {author} {\bibfnamefont {E.}~\bibnamefont
  {Sch\"oll}},\ }\href {\doibase 10.1103/PhysRevLett.110.224101} {\bibfield
  {journal} {\bibinfo  {journal} {Phys. Rev. Lett.}\ }\textbf {\bibinfo
  {volume} {110}},\ \bibinfo {pages} {224101} (\bibinfo {year}
  {2013})}\BibitemShut {NoStop}%
\bibitem [{\citenamefont {Omel'chenko}(2013)}]{Nonlinearity.26.2469}%
  \BibitemOpen
  \bibfield  {author} {\bibinfo {author} {\bibfnamefont {O.~E.}\ \bibnamefont
  {Omel'chenko}},\ }\href {http://stacks.iop.org/0951-7715/26/i=9/a=2469}
  {\bibfield  {journal} {\bibinfo  {journal} {Nonlinearity}\ }\textbf {\bibinfo
  {volume} {26}},\ \bibinfo {pages} {2469} (\bibinfo {year}
  {2013})}\BibitemShut {NoStop}%
\bibitem [{\citenamefont {Schmidt}\ \emph {et~al.}(2014)\citenamefont
  {Schmidt}, \citenamefont {Sch\"onleber}, \citenamefont {Krischer},\ and\
  \citenamefont {Garc\'ia-Morales}}]{Chaos.24.013102}%
  \BibitemOpen
  \bibfield  {author} {\bibinfo {author} {\bibfnamefont {L.}~\bibnamefont
  {Schmidt}}, \bibinfo {author} {\bibfnamefont {K.}~\bibnamefont
  {Sch\"onleber}}, \bibinfo {author} {\bibfnamefont {K.}~\bibnamefont
  {Krischer}}, \ and\ \bibinfo {author} {\bibfnamefont {V.}~\bibnamefont
  {Garc\'ia-Morales}},\ }\href {\doibase 10.1063/1.4858996} {\bibfield
  {journal} {\bibinfo  {journal} {Chaos}\ }\textbf {\bibinfo {volume} {24}},\
  \bibinfo {pages} {013102} (\bibinfo {year} {2014})}\BibitemShut {NoStop}%
\bibitem [{\citenamefont {Maistrenko}\ \emph {et~al.}(2014)\citenamefont
  {Maistrenko}, \citenamefont {Vasylenko}, \citenamefont {Sudakov},
  \citenamefont {Levchenko},\ and\ \citenamefont
  {Maistrenko}}]{IJBC.24.1440014}%
  \BibitemOpen
  \bibfield  {author} {\bibinfo {author} {\bibfnamefont {Y.~L.}\ \bibnamefont
  {Maistrenko}}, \bibinfo {author} {\bibfnamefont {A.}~\bibnamefont
  {Vasylenko}}, \bibinfo {author} {\bibfnamefont {O.}~\bibnamefont {Sudakov}},
  \bibinfo {author} {\bibfnamefont {R.}~\bibnamefont {Levchenko}}, \ and\
  \bibinfo {author} {\bibfnamefont {V.~L.}\ \bibnamefont {Maistrenko}},\ }\href
  {\doibase 10.1142/S0218127414400148} {\bibfield  {journal} {\bibinfo
  {journal} {Int. J. Bifurc. Chaos}\ }\textbf {\bibinfo {volume} {24}},\
  \bibinfo {pages} {1440014} (\bibinfo {year} {2014})}\BibitemShut {NoStop}%
\bibitem [{\citenamefont {Xie}\ \emph {et~al.}(2014)\citenamefont {Xie},
  \citenamefont {Knobloch},\ and\ \citenamefont {Kao}}]{PhysRevE.90.022919}%
  \BibitemOpen
  \bibfield  {author} {\bibinfo {author} {\bibfnamefont {J.}~\bibnamefont
  {Xie}}, \bibinfo {author} {\bibfnamefont {E.}~\bibnamefont {Knobloch}}, \
  and\ \bibinfo {author} {\bibfnamefont {H.-C.}\ \bibnamefont {Kao}},\ }\href
  {\doibase 10.1103/PhysRevE.90.022919} {\bibfield  {journal} {\bibinfo
  {journal} {Phys. Rev. E}\ }\textbf {\bibinfo {volume} {90}},\ \bibinfo
  {pages} {022919} (\bibinfo {year} {2014})}\BibitemShut {NoStop}%
\bibitem [{\citenamefont {Rosin}\ \emph {et~al.}(2014)\citenamefont {Rosin},
  \citenamefont {Rontani}, \citenamefont {Haynes}, \citenamefont {Sch\"oll},\
  and\ \citenamefont {Gauthier}}]{PhysRevE.90.030902}%
  \BibitemOpen
  \bibfield  {author} {\bibinfo {author} {\bibfnamefont {D.~P.}\ \bibnamefont
  {Rosin}}, \bibinfo {author} {\bibfnamefont {D.}~\bibnamefont {Rontani}},
  \bibinfo {author} {\bibfnamefont {N.~D.}\ \bibnamefont {Haynes}}, \bibinfo
  {author} {\bibfnamefont {E.}~\bibnamefont {Sch\"oll}}, \ and\ \bibinfo
  {author} {\bibfnamefont {D.~J.}\ \bibnamefont {Gauthier}},\ }\href {\doibase
  10.1103/PhysRevE.90.030902} {\bibfield  {journal} {\bibinfo  {journal} {Phys.
  Rev. E}\ }\textbf {\bibinfo {volume} {90}},\ \bibinfo {pages} {030902}
  (\bibinfo {year} {2014})}\BibitemShut {NoStop}%
\bibitem [{\citenamefont {Ashwin}\ and\ \citenamefont
  {Burylko}(2015)}]{Chaos.25.013106}%
  \BibitemOpen
  \bibfield  {author} {\bibinfo {author} {\bibfnamefont {P.}~\bibnamefont
  {Ashwin}}\ and\ \bibinfo {author} {\bibfnamefont {O.}~\bibnamefont
  {Burylko}},\ }\href {\doibase 10.1063/1.4905197} {\bibfield  {journal}
  {\bibinfo  {journal} {Chaos}\ }\textbf {\bibinfo {volume} {25}},\ \bibinfo
  {pages} {013106} (\bibinfo {year} {2015})}\BibitemShut {NoStop}%
\bibitem [{\citenamefont {Panaggio}\ and\ \citenamefont
  {Abrams}(2015)}]{Nonlinearity.28.R67}%
  \BibitemOpen
  \bibfield  {author} {\bibinfo {author} {\bibfnamefont {M.~J.}\ \bibnamefont
  {Panaggio}}\ and\ \bibinfo {author} {\bibfnamefont {D.~M.}\ \bibnamefont
  {Abrams}},\ }\href {http://stacks.iop.org/0951-7715/28/i=3/a=R67} {\bibfield
  {journal} {\bibinfo  {journal} {Nonlinearity}\ }\textbf {\bibinfo {volume}
  {28}},\ \bibinfo {pages} {R67} (\bibinfo {year} {2015})}\BibitemShut
  {NoStop}%
\bibitem [{\citenamefont {Haugland}\ \emph {et~al.}(2015)\citenamefont
  {Haugland}, \citenamefont {Schmidt},\ and\ \citenamefont
  {Krischer}}]{SciRep.5.9883}%
  \BibitemOpen
  \bibfield  {author} {\bibinfo {author} {\bibfnamefont {S.~W.}\ \bibnamefont
  {Haugland}}, \bibinfo {author} {\bibfnamefont {L.}~\bibnamefont {Schmidt}}, \
  and\ \bibinfo {author} {\bibfnamefont {K.}~\bibnamefont {Krischer}},\ }\href
  {\doibase 10.1038/srep09883} {\bibfield  {journal} {\bibinfo  {journal} {Sci.
  Rep.}\ }\textbf {\bibinfo {volume} {5}},\ \bibinfo {pages} {9883} (\bibinfo
  {year} {2015})}\BibitemShut {NoStop}%
\bibitem [{\citenamefont {Schmidt}\ and\ \citenamefont
  {Krischer}(2015)}]{Chaos.25.064401}%
  \BibitemOpen
  \bibfield  {author} {\bibinfo {author} {\bibfnamefont {L.}~\bibnamefont
  {Schmidt}}\ and\ \bibinfo {author} {\bibfnamefont {K.}~\bibnamefont
  {Krischer}},\ }\href {\doibase 10.1063/1.4921727} {\bibfield  {journal}
  {\bibinfo  {journal} {Chaos}\ }\textbf {\bibinfo {volume} {25}},\ \bibinfo
  {pages} {064401} (\bibinfo {year} {2015})}\BibitemShut {NoStop}%
\bibitem [{\citenamefont {Omelchenko}\ \emph {et~al.}(2015)\citenamefont
  {Omelchenko}, \citenamefont {Zakharova}, \citenamefont {H\"ovel},
  \citenamefont {Siebert},\ and\ \citenamefont {Sch\"oll}}]{Chaos.25.083104}%
  \BibitemOpen
  \bibfield  {author} {\bibinfo {author} {\bibfnamefont {I.}~\bibnamefont
  {Omelchenko}}, \bibinfo {author} {\bibfnamefont {A.}~\bibnamefont
  {Zakharova}}, \bibinfo {author} {\bibfnamefont {P.}~\bibnamefont {H\"ovel}},
  \bibinfo {author} {\bibfnamefont {J.}~\bibnamefont {Siebert}}, \ and\
  \bibinfo {author} {\bibfnamefont {E.}~\bibnamefont {Sch\"oll}},\ }\href
  {\doibase 10.1063/1.4927829} {\bibfield  {journal} {\bibinfo  {journal}
  {Chaos}\ }\textbf {\bibinfo {volume} {25}},\ \bibinfo {pages} {083104}
  (\bibinfo {year} {2015})}\BibitemShut {NoStop}%
\bibitem [{\citenamefont {Suda}\ and\ \citenamefont
  {Okuda}(2015)}]{PhysRevE.92.060901}%
  \BibitemOpen
  \bibfield  {author} {\bibinfo {author} {\bibfnamefont {Y.}~\bibnamefont
  {Suda}}\ and\ \bibinfo {author} {\bibfnamefont {K.}~\bibnamefont {Okuda}},\
  }\href {\doibase 10.1103/PhysRevE.92.060901} {\bibfield  {journal} {\bibinfo
  {journal} {Phys. Rev. E}\ }\textbf {\bibinfo {volume} {92}},\ \bibinfo
  {pages} {060901} (\bibinfo {year} {2015})}\BibitemShut {NoStop}%
\bibitem [{\citenamefont {Panaggio}\ \emph {et~al.}(2016)\citenamefont
  {Panaggio}, \citenamefont {Abrams}, \citenamefont {Ashwin},\ and\
  \citenamefont {Laing}}]{PhysRevE.93.012218}%
  \BibitemOpen
  \bibfield  {author} {\bibinfo {author} {\bibfnamefont {M.~J.}\ \bibnamefont
  {Panaggio}}, \bibinfo {author} {\bibfnamefont {D.~M.}\ \bibnamefont
  {Abrams}}, \bibinfo {author} {\bibfnamefont {P.}~\bibnamefont {Ashwin}}, \
  and\ \bibinfo {author} {\bibfnamefont {C.~R.}\ \bibnamefont {Laing}},\ }\href
  {\doibase 10.1103/PhysRevE.93.012218} {\bibfield  {journal} {\bibinfo
  {journal} {Phys. Rev. E}\ }\textbf {\bibinfo {volume} {93}},\ \bibinfo
  {pages} {012218} (\bibinfo {year} {2016})}\BibitemShut {NoStop}%
\bibitem [{\citenamefont {Smirnov}\ \emph {et~al.}(2017)\citenamefont
  {Smirnov}, \citenamefont {Osipov},\ and\ \citenamefont
  {Pikovsky}}]{JPhysA.50.08LT01}%
  \BibitemOpen
  \bibfield  {author} {\bibinfo {author} {\bibfnamefont {L.}~\bibnamefont
  {Smirnov}}, \bibinfo {author} {\bibfnamefont {G.}~\bibnamefont {Osipov}}, \
  and\ \bibinfo {author} {\bibfnamefont {A.}~\bibnamefont {Pikovsky}},\ }\href
  {http://stacks.iop.org/1751-8121/50/i=8/a=08LT01} {\bibfield  {journal}
  {\bibinfo  {journal} {J. Phys. A}\ }\textbf {\bibinfo {volume} {50}},\
  \bibinfo {pages} {08LT01} (\bibinfo {year} {2017})}\BibitemShut {NoStop}%
\bibitem [{\citenamefont {Cheng}\ \emph {et~al.}(2018)\citenamefont {Cheng},
  \citenamefont {Dai}, \citenamefont {Wu}, \citenamefont {Feng}, \citenamefont
  {Li},\ and\ \citenamefont {Yang}}]{CNSNS.56.1}%
  \BibitemOpen
  \bibfield  {author} {\bibinfo {author} {\bibfnamefont {H.}~\bibnamefont
  {Cheng}}, \bibinfo {author} {\bibfnamefont {Q.}~\bibnamefont {Dai}}, \bibinfo
  {author} {\bibfnamefont {N.}~\bibnamefont {Wu}}, \bibinfo {author}
  {\bibfnamefont {Y.}~\bibnamefont {Feng}}, \bibinfo {author} {\bibfnamefont
  {H.}~\bibnamefont {Li}}, \ and\ \bibinfo {author} {\bibfnamefont
  {J.}~\bibnamefont {Yang}},\ }\href {\doibase 10.1016/j.cnsns.2017.07.015}
  {\bibfield  {journal} {\bibinfo  {journal} {Commun. Nonlinear Sci. Numer.
  Simul.}\ }\textbf {\bibinfo {volume} {56}},\ \bibinfo {pages} {1} (\bibinfo
  {year} {2018})}\BibitemShut {NoStop}%
\bibitem [{\citenamefont {Sakaguchi}\ and\ \citenamefont
  {Kuramoto}(1986)}]{ProgTheorPhys.76.576}%
  \BibitemOpen
  \bibfield  {author} {\bibinfo {author} {\bibfnamefont {H.}~\bibnamefont
  {Sakaguchi}}\ and\ \bibinfo {author} {\bibfnamefont {Y.}~\bibnamefont
  {Kuramoto}},\ }\href {\doibase 10.1143/PTP.76.576} {\bibfield  {journal}
  {\bibinfo  {journal} {Prog. Theor. Phys.}\ }\textbf {\bibinfo {volume}
  {76}},\ \bibinfo {pages} {576} (\bibinfo {year} {1986})}\BibitemShut
  {NoStop}%
\bibitem [{\citenamefont {Watanabe}\ and\ \citenamefont
  {Strogatz}(1994)}]{PhysicaD.74.197}%
  \BibitemOpen
  \bibfield  {author} {\bibinfo {author} {\bibfnamefont {S.}~\bibnamefont
  {Watanabe}}\ and\ \bibinfo {author} {\bibfnamefont {S.~H.}\ \bibnamefont
  {Strogatz}},\ }\href {\doibase https://doi.org/10.1016/0167-2789(94)90196-1}
  {\bibfield  {journal} {\bibinfo  {journal} {Physica D}\ }\textbf {\bibinfo
  {volume} {74}},\ \bibinfo {pages} {197 } (\bibinfo {year}
  {1994})}\BibitemShut {NoStop}%
\bibitem [{\citenamefont {Ott}\ and\ \citenamefont
  {Antonsen}(2008)}]{Chaos.18.037113}%
  \BibitemOpen
  \bibfield  {author} {\bibinfo {author} {\bibfnamefont {E.}~\bibnamefont
  {Ott}}\ and\ \bibinfo {author} {\bibfnamefont {T.~M.}\ \bibnamefont
  {Antonsen}},\ }\href {\doibase 10.1063/1.2930766} {\bibfield  {journal}
  {\bibinfo  {journal} {Chaos}\ }\textbf {\bibinfo {volume} {18}},\ \bibinfo
  {pages} {037113} (\bibinfo {year} {2008})}\BibitemShut {NoStop}%
\bibitem [{\citenamefont {Ott}\ and\ \citenamefont
  {Antonsen}(2009)}]{Chaos.19.023117}%
  \BibitemOpen
  \bibfield  {author} {\bibinfo {author} {\bibfnamefont {E.}~\bibnamefont
  {Ott}}\ and\ \bibinfo {author} {\bibfnamefont {T.~M.}\ \bibnamefont
  {Antonsen}},\ }\href {\doibase 10.1063/1.3136851} {\bibfield  {journal}
  {\bibinfo  {journal} {Chaos}\ }\textbf {\bibinfo {volume} {19}},\ \bibinfo
  {pages} {023117} (\bibinfo {year} {2009})}\BibitemShut {NoStop}%
\bibitem [{\citenamefont {Zhang}\ \emph {et~al.}(2017)\citenamefont {Zhang},
  \citenamefont {Pikovsky},\ and\ \citenamefont {Liu}}]{SciRep.7.2104}%
  \BibitemOpen
  \bibfield  {author} {\bibinfo {author} {\bibfnamefont {X.}~\bibnamefont
  {Zhang}}, \bibinfo {author} {\bibfnamefont {A.}~\bibnamefont {Pikovsky}}, \
  and\ \bibinfo {author} {\bibfnamefont {Z.}~\bibnamefont {Liu}},\ }\href
  {\doibase 10.1038/s41598-017-02283-1} {\bibfield  {journal} {\bibinfo
  {journal} {Sci. Rep.}\ }\textbf {\bibinfo {volume} {7}},\ \bibinfo {pages}
  {2104} (\bibinfo {year} {2017})}\BibitemShut {NoStop}%
\end{thebibliography}%

\end{document}